\documentstyle[epsf]{mn}
\def\gsim{\;\lower.6ex\hbox{$\sim$}\kern-7.75pt\raise.65ex\hbox{$>$}\;}
\def\lsim{\;\lower.6ex\hbox{$\sim$}\kern-7.75pt\raise.65ex\hbox{$<$}\;}

\title[On the Existence of Differences in Luminosity between HB stars]
{On the 
Existence of Differences in Luminosity between Horizontal Branch Stars
in Globular Clusters and in the Field}

\author[Carretta et al.]{E. Carretta$^1$, R.G. Gratton$^1$, G. Clementini$^2$\\ 
 $^1$ Osservatorio Astronomico di Padova, Italy, 
      e-mail carretta@pd.astro.it, gratton@pd.astro.it\\
 $^2$ Osservatorio Astronomico di Bologna, Italy, 
      e-mail gisella@astbo3.bo.astro.it}
\date{}

\begin{document}
\maketitle

\begin{abstract}
The discrepancy between a {\it long} distance scale derived from 
Hipparcos based distances to globular clusters via main sequence fitting 
to local subdwarfs,
and a {\it short} distance scale derived from the absolute magnitude of 
field RR Lyraes via statistical parallaxes and the Baade-Wesselink method
could be accounted for whether an intrinsic difference 
of about $\sim$ 0.1-0.2 mag was found to exist between horizontal 
branch (HB) stars 
populating the {\it sparse} general field and the {\it dense} globular
clusters. In this paper we discuss the possible existence of such a systematic 
difference comparing the {\it period-shifts} observed for field and cluster 
RR Lyraes.
 Various approaches based on different parameters and data-sets for both 
cluster and field variables were used in order to establish the size 
of such a hypothetical
difference, if any. We find that on the whole very small not significant 
differences
exist between the period-metallicity distributions of field and cluster RR 
Lyraes, thus confirming  with a more quantitative approach, the 
qualitative conclusions by Catelan (1998). 
This observational 
evidence translates into a very small difference
between the horizontal branch luminosity of field and cluster stars, 
unless RR Lyraes in Globular Clusters are about 0.06 
M$_\odot$ more
massive than field RR Lyrae at same metallicity, which is to be proven.
\end{abstract}


\section{Introduction}

The well-known dichotomy existing between {\it short} and {\it long} distance 
scale
as derived from old, Population II stars is still an unsolved problem, 
not yet
settled even after the improvement, in distance 
determinations, 
due to the release of the Hipparcos Parallax Catalogue.
In fact, while distances to globular clusters by main sequence fitting 
to local subdwarfs with parallaxes measured by Hipparcos, 
favour the {\it long} distance scale 
 (see Gratton et al. 1997, Paper I, and Carretta et al. 2000, Paper II, 
for extensive updates and 
discussions on this topic), the statistical parallaxes of field RR Lyraes 
(one of 
the most commonly used 
galactic {\it standard candle}),
based on Hipparcos proper motions (Fernley et al. 1998; Popowski and Gould,
1998, Tsujimoto et al., 1997) 
still lead to the {\it short} distance scale.

Following an alternative approach, Gratton (1998) used Hipparcos parallaxes 
for a sample of field metal-poor HB stars in order to directly calibrate these 
standard candles.
Given the paucity of RR Lyrae variables within reasonable 
distances from the Sun, Hipparcos was able to measure useful parallaxes
for only 3 variables, RR Lyrae itself and two additional stars. Uncertainties 
in the parallax determinations of 
the latter are however rather large.
In order to increase the 
statistical significance of the sample, Gratton (1998)  selected also 
red and blue HB field
stars from various sources. His final sample consists of 20 stars with $V< 9$
and 2 stars slightly fainter than this limit. For a consistent 
comparison of the results, metal abundances for the stars in the sample 
were put on the 
same metallicity scale used 
in Paper I and II.
The mean weighted absolute magnitude found by 
Gratton (1998) with this procedure is $M_V=+0.69\pm 0.10$
(at average metallicity [Fe/H]=$-1.41$), and brightens to
$M_V=+0.60\pm 0.12$ (at average [Fe/H]=$-1.51$) when HD17072, 
a suspected first ascent giant branch star (see however Carney, Lee \& 
Habgood, 1998) is discarded from the sample.
This latter value has been recently revised to $M_V=+0.62\pm 0.11$ 
(see Koen \& Laney 1998) in order to account for the intrinsic scatter 
in the HB 
magnitudes when correcting for the Lutz-Kelker effect.
The error bar given by this analysis is still large enough that a final choice
between the short and long distance scale can not be made.

Gratton (1998), however pointed
out how different distance determination methods seem to give consistent 
results as far as   
only $field$ HB stars or only 
{\it globular cluster} HB stars are considered.
This argument led him to suggest that a real difference in luminosity 
of $\sim 0.1\div 0.2$ mag, 
might actually
exist between HB stars in globular cluster and in the general
field, the cluster stars being brighter.

The hypothesis of an intrinsic difference between field and
cluster RR Lyraes was immediately challenged by 
Catelan (1998; C98) who used the period-temperature
distribution for both field and cluster variables at about fixed 
metallicity (in 
a metal-poor and a metal-rich regime) to show that 
GC and field RR Lyrae are essentially indistinguishable in 
the $P - T_{eq}$ plane, thus concluding that there is no significant difference
in luminosity between them.

In the present paper we try to have a deeper insight into this problem.
Both the original targets in C98 analysis as well as a new, more 
homogeneous sample of cluster and field variables are used in order 
to achieve a {\it quantitative} estimate of the possible
differences in luminosity between HB field and cluster stars, 
and to complement and
refine the fully qualitative approach used by C98.

\section{A reanalysis of C98 data}

As a starting point, we have re-analyzed the original set of data considered
by C98, kindly provided by the author. C98 dataset consists of 
35 field RR Lyraes and 49 variables in 5 globular clusters (namely NGC 
362, M5, M68, M15 and M92). 
Variables were selected by C98 in order to be of ab-type, and with
light curves not affected by Blazhko effect. Metallicities were from
Layden et al. (1996), and are therefore on Zinn \& West (1984) metallicity 
scale, which is 
also adopted by C98 for the globular cluster variables.

According to C98 we used the empirical relation of 
Carney, Storm \& Jones (1992a; CSJ) which gives the so-called ``equilibrium
temperature" T$_{\rm eq}$ of a RR Lyrae variable as a function of the blue
amplitude $A_B$, metal abundance [Fe/H] and pulsational period P.
This relation  
provides a very tight logP--logT$_{\rm eq}$ between cluster and field variables
(see figure 2 in C98), and hence seems to be the best confirmation
of a similarity in their luminosities.

\begin{figure}
\vspace{8cm}
\includegraphics{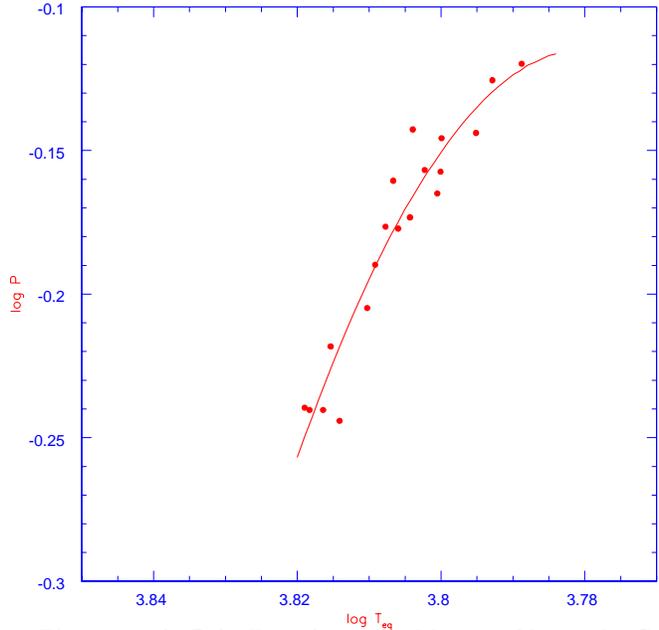}
\caption{ logP--logT$_{\rm eq}$ relation for M15 variables in the C98
list. Equilibrium temperatures are obtained using the empirical relation of
CSJ (their eq. 16) and P,$A_B$ and [Fe/H] values selected by C98.} 
\label{fig1} 
\end{figure}

Following an approach similar to that of C98, we then derived T$_{\rm eq}$
values for all stars in his sample. However, in order to quantify
the supposed identity between field and cluster
variables at fixed metal abundance we used the M15 variables as a reference. 
A quadratic relationship was fit to the M15 RR Lyraes in
the logP--logT$_{\rm eq}$ plane (see Figure~\ref{fig1}), and we derived 
for each variable  
the expected period if the star was to follow the 
M15 logP--logT$_{\rm eq}$ relation.

Differences
between the actually observed (as quoted by C98) and
the expected period were then computed for each star : 
$\Delta$Ps $=$ dlogP$_{\rm oss - exp}$.

\begin{figure}
\vspace{8cm}
\includegraphics{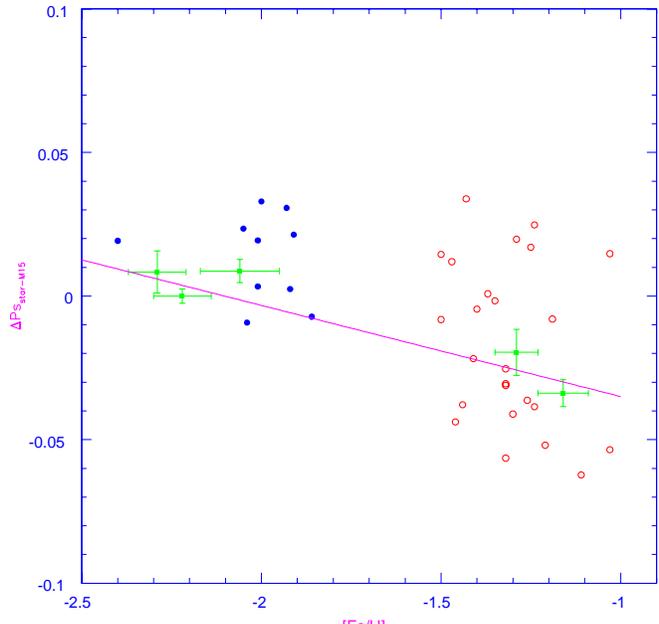}
\caption{ Run of the $\Delta$Ps $=$ dlogP$_{\rm oss - exp}$ as a function of
the metal abundance for stars in C98 original sample.
Filled and open circles are stars defined respectively metal-poor and
metal-rich by C98, while filled squares represent the average values for
cluster variables. For the latter, error bars represent internal errors. The
straight line is the linear best-fit to the cluster data. Metallicities are
on the Layden--Zinn \& West scale (Layden et al. 1996).}
\label{fig2} 
\end{figure}

Figure~\ref{fig2} shows the run of $\Delta$Ps $vs$ [Fe/H] for all stars in the
original C98 sample. For the cluster variables we show the average value
at the cluster metallicity. 

A linear best-fit was drawn through the cluster data 
in order to evaluate how much field stars differ
from the cluster variables. 
Admittedly, there is some hint that fitting by a second order polynomial 
could be a better representation of the data, however the lack of
clusters around [Fe/H]$= -1.5 \div -1.8$ in C98 
original sample prevents further assessment of this issue. 
We then computed the distance 
of each field star from the fit representing the period-shift $vs$ [Fe/H]
relation defined by the cluster variables. 

The unweighted average derived from all 35 (both metal-poor and metal-rich) 
field stars is :
\begin{eqnarray}
<\delta(\Delta Ps)> & = & <\delta(dlogP_{star} - dlogP_{fit,GC})> \nonumber \\
                    & = & +0.0108\pm 0.0042
\end{eqnarray}

\noindent
with $\sigma=0.0246$. 
This relation is a more quantitative measure of the results
achieved by C98,  
and simply tells us that the field variables 
have at least the same, or possibly even longer period 
shifts (within 
rather large internal errors) with respect to cluster variables of comparable
metal abundance. This could be interpreted as a (weak) evidence for 
the field RR Lyraes to be $brighter$ than
their cluster counterparts. A result a little unpalatable, since the
discrepancy between distance scales could be accounted for only whether 
field RR Lyraes were {\it fainter} than their cluster counterparts.
We should recall, however, that what the period shift between the two
distributions (at fixed temperature and metallicity) actually measures is the
the convolution of stellar mass $and$ luminosity effects, as in
the classical equation of pulsation by van Albada \& Baker (1971) (e.g. eq. 10
in CSJ or the more recent one by Bono et al. 1997). This point will 
be further discussed below.

\subsection{The effect of changing dataset}

Having settled the size of the effect we can expect, and of the related 
errors, we then explored how the observational 
parameters involved in the analysis could affect the derived result.

We thus repeated the analysis adopting different 
sets of parameters, metallicity scale (from Clementini et
al. 1995, C95 or Blanco 1992 for field variables; and from Carretta \& Gratton 
1997, CG97, or Zinn \& West 1984 for clusters) and light curve parameters
(as the updated blue amplitudes for field stars by Nikolov et al. 1984).

These tests allowed us to ascertain that:
\begin{enumerate}
\item The effect that we want to highlight is (or could be) very subtle: in
the best case we want to detect a difference of about 0.1-0.15 mag in 
the HB luminosities of cluster and field stars.
\item The internal errors alone are very likely about the same size of the 
effect we are looking for. Even parameters usually thought to be very reliable
and simple to measure, as amplitudes and periods, should be carefully 
checked.
\item There is a strong suggestion that homogeneous data sets could greatly
help to reduce systematic errors which may smear out real differences.
\end{enumerate}

\subsection{Is the pulsational approach the proper way to detect
possible luminosity differences between field and cluster stars ?}

There may be some procedural concerns about the use of the pulsational 
approach to detect a systematic difference in
luminosity between field and cluster HB stars.

A first concern was pointed out by Catelan (1998), who suggested that 
an
empirical calibration of T$_{\rm eq}$ omitting the period term would be
a better approach, with respect to the original CSJ calibration.
In fact, there is a risk of being catched in a circular line of thought, 
entering with pulsational periods in order to derive temperatures and then
using distributions of field and cluster stars in the T$_{\rm eq} -$logP 
plane to highlight a difference (shift) in the periods, to be interpreted as
a luminosity difference.
This can be seen also by a simple experiment suggested by the referee. If we
decrease by 0.2 mag the luminosities of variables in Figure 1 in order to
simulate a fainter sample, then the combination of the van Albada \& Baker
pulsation equation and of eq. (16) in CSJ acts to shift these
fainter variables toward shorter periods and higher temperatures. The net effect
is to transport the overall distribution in the T$_{\rm eq}$ -logP plane
along the relationship (and its extension to higher temperatures), so that no
significant period shift can be detected between the original, observed sample
of M 15 variables and the simulated faint one.

Unfortunately, the relationship for T$_{\rm eq}$ derived by C98 
with no period term 
is so scattered (see Figure 3 in C98), that no useful information can be 
derived, apart from a generic similarity of the field and cluster star 
distributions in the T$_{\rm eq} -$logP plane. 
Moreover, a reference 
relationship logP $vs$ log T$_{\rm eq}$ seems not very easy to establish for
any of the clusters in C98 list, judging from his Figure 3.

A point of further concern is that 
both in the
original CSJ and in C98 analysis, 
the empirical calibrations
of T$_{\rm eq}$ are derived using only the field stars and then 
applied also to the cluster
variables. The underlying assumption is that RR Lyrae stars share the
same physical parameters, with no dependence on the environment
in which they were born : the dense clusters or the much looser field.
We believe that this is a dangerous procedure which may 
mimic spurious and/or mask real differences in the absolute
magnitude of field and cluster variables.

\section{An independent evaluation of C98 results}

In 
order to quantitatively assess the analysis by C98  
we need a more homogeneous sample of stars and a
temperature calibration which can be  applied independently to cluster and 
field RR Lyraes.
We have thus selected a sample of field stars 
smaller in numerical size with respect to C98, but which can provide a 
tighter distribution, thanks to the higher degree of accuracy 
of their photometric data.
Using new model atmospheres by Kurucz (1993) and semi-empirical 
colour-temperature calibrations, 
new temperatures were derived for all field and cluster variables, irrespective
 of their belonging to one or the other enviroment.
Selected samples and their analysis are  discussed in the following section.

\subsection{Definition of a new sample, observables and derived temperatures}

Our new sample of field stars consists of 16 ab-type RR Lyraes used by CSJ
in Baade-Wesselink analyses (we disregarded RS Boo since it is presently known
to be affected by Blazhko).
The new targets satisfy all requisites 
listed in section 3.1 of C98 : being the variables usually used in the
Baade-Wesselink approach to distances, they all have 
a very good coverage at
each phase of light curve, periods and amplitudes are measured with much greater accuracy
than for variables in generic surveys as those of Layden et al. (1996) and all 
have new homogeneous metallicity estimates from the new metallicity scale
for the RR Lyraes variables of C95\footnote{The metal abundance scale
in C95 is fully consistent with the metallicity scale derived by CG97 for globular 
cluster giants (see discussion in CG97)}.

As in CSJ, we have not eliminated the supposedly evolved stars
DX Del and SS Leo, since we are comparing sets of stars (field and
clusters') which both are likely to include some evolved variables.

As for the cluster variables, we considered 19 RRab in M15 (the same used 
by C98), 41 RRab variables in M 3 with new CCD photometry from Carretta et 
al. (1998), 22 RRab in M 5 (Ripepi et al. 1998, private communication),
8 type-ab RR Lyrae with clean light curves in M 68 (Walker 1994), 6 RRab
in M 92 (Carney et al. 1992b) and 18 RRab in NGC 6362 (Walker 1999, 
private communication).
All observational datasets but those for M15 are
from recent CCD $B,V$ photometry.  
However, we are confident that even
if based on photographic observations, derived quantities for M15 variables 
well compare with the CCD data for the other clusters. In fact,  
 none of our conclusions would be altered whether 
 any other of the clusters in our sample was taken as a reference, even if we
would have used NGC 6362, whose stars span all the relevant range in 
temperature. 
In particular, the period shift relations (see next section) are not
sensitive to the choice of the cluster used to define the
fiducial locus for determining the period shifts.

 We used the magnitude-average color index $(B-V)_{mag}$ as temperature  
indicator. As widely discussed in CSJ, there
is no particular reason to believe the B$-$V  to be the best colour index to
reproduce the RR Lyrae's equilibrium temperature. 
However, B,V photometry is presently available for a larger number of cluster 
variables, and, 
on the other hand, 
since 
this is a $differential$ analysis between cluster and field stars, 
the crucial point is to use 
the same colour (no matter which), for both kind of variables, in 
order to achieve the maximum
degree of homogeneity.

$<B>-<V>$ colours (where  
$<B>$ and $<V>$ are intensity-averaged magnitudes), are available for all 
cluster variables, as
well as for the field sample adopted by CSJ (column 6 of their Table 4).

Using the latter we derived 
a relation to transform the intensity-averaged $<B>-<V>$ colours into 
magnitude-averaged $(B-V)_{mag}$ colours for all variables in our sample.

\begin{figure}
\vspace{8cm}
\includegraphics{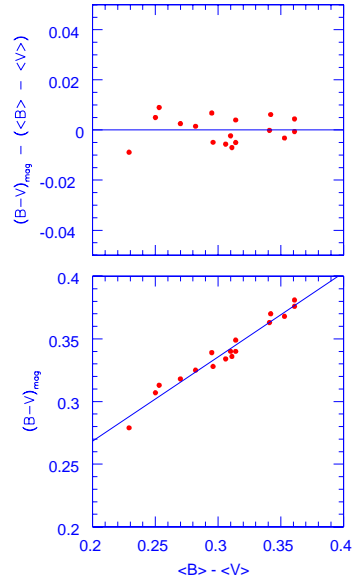}
\caption{ Conversion of the intensity-averaged $<B>-<V>$ in the magnitude-averaged 
$(B-V)_{mag}$ colour for the field variables in the Baade-Wesselink sample of
Carney et al. (1992a).}
\label{fig5} 
\end{figure}

Figure~\ref{fig5} displays this colour conversion derived from the field
Baade-Wesselink stars of CSJ (column 6 of their Table 4). The best-fitting 
relation is :
\begin{eqnarray}
(B-V)_{mag} - (<B>-<V>) = \nonumber \\
-0.328(\pm 0.035) \cdot (<B>-<V>) +0.134(\pm 0.001) 
\end{eqnarray}
$r.m.s.$ = 0.006, correlation coefficient $r=0.92$, based
on 17 stars. 

Although the two indices are not in a 
one-to-one correspondence, the above transformation allows us to obtain a 
set of colours mutually consistent for both the
field and the cluster stars in our samples.

Temperatures were obtained for all stars using this colour and the 
semi-empirical procedure discussed at length in Gratton, Carretta \& Castelli
(1996). 
Briefly, the latest model atmospheres by Kurucz (1993) were tied (i.e. 
``corrected with") to an empirical colour-temperature calibration for pop. I
stars based on the Infrared Flux Method (see details in Gratton, Carretta \&
Castelli). The corrected models then were used to read from the observed 
de-reddened colours the temperatures at the gravity and metal abundance 
appropriate for each star.

A constant value of $\log g=2.75$ was adopted for all variables. 
This is the value generally adopted for 
RR Lyraes at minimum light and presently there is some indication 
that a larger value by 0.10-0.15 dex might be more appropriate for RR Lyraes
at minimum light. However the results of our differential 
analysis are not affected by the adopted value  
of $\log g$ provided that the same value is adopted  
for both field and cluster 
variables\footnote{Strictly speaking, gravity should be slighthly 
different in RR Lyraes
of different luminosities and masses. However, changing gravity by 0.1 dex
(corresponding to a difference of 25\% in the mass and/or 0.25 mag in
magnitude) only changes T$_{\rm eff}$'s by 10 K. The effect on the period
shifts is only of 0.002 dex and can be neglected when compared with
uncertainties e.g. on interstellar reddening}. 
Metal abundances for the field stars were from
C95. Reddening for M15 ($E(B-V)=0.09\pm 0.01)$, M3 ($E(B-V)=0.02\pm 0.01)$,
M5 ($E(B-V)=0.035\pm 0.005)$,  M92 ($E(B-V)=0.025\pm 0.005)$ and M 68 
($E(B-V)=0.04\pm 0.01$) are fully consistent with
the new reddening scale of Paper I (Gratton et al. 1997) and Paper II 
(Carretta et al. 2000). In turn,
metal abundances for the cluster variables were slightly different from the
values of the original CG97 scale. Taking into account the adopted reddenings,
the temperatures for the cluster RR Lyraes were thus obtained using [Fe/H]$=-1.30$
for M 3, [Fe/H]$=-2.14$ for M15, [Fe/H]$=-1.95$ for M 68, [Fe/H]$=-1.10$
for M 5 and [Fe/H]$=-2.15$ for M 92. For NGC 6362 (not included in the 
sample of Paper I and II), we adopt [Fe/H]=$-$0.96 dex from Carretta \& Gratton 
(1997)
and E(B$-$V)=0.08 mag from Brocato et al. (1999).

\subsection{Analysis}

The only assumption made so far is that for each variable,
either in the field or in a cluster, a temperature can be defined using the 
latest
Kurucz model atmospheres (empirically corrected) and that the
temperatures so derived represent the ones the variables
would have if they were static stars. Bearing in mind that we aim at a 
differential comparison, we may ask how much this assumption can be
trusted.
 
 As an estimate we can compare our newly derived temperatures with 
the equilibrium temperatures for field stars
analyzed with the Baade-Wesselink method, listed in column 9 of Table 4 in
CSJ.

\begin{figure}
\vspace{8cm}
\includegraphics{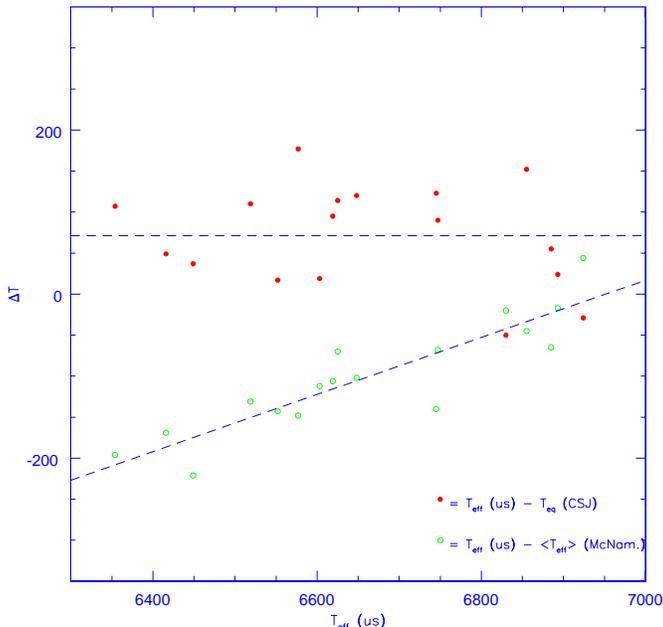}
\caption{ Temperature differences for 17 stars with Baade-Wesselink analysis
as a function of the newly derived T$_{eff}$. Differences 
{\it this 
paper-other studies} 
are with the 
T$_{\rm eq}$ of CSJ (filled circles) and the $<T_{eff}>$ values of 
McNamara (1997; open symbols).}
\label{fig6} 
\end{figure}

This comparison is shown in Figure~\ref{fig6}.
Our temperatures are on average larger than the equilibrium temperatures of
CSJ, the mean difference being 
$<T_{\rm eff, this paper} - T_{\rm eq, CSJ}> = 71 \pm 15$~K ($r.m.s.= 61$~K, 17 
stars),
with no trend with temperature (or metal-abundance).

Taken at face value, this indicates that equilibrium temperatures for
pulsating pop. II stars as the RR Lyraes, computed from infrared colours 
(as those
usually employed in Baade-Wesselink analyses) and the old Kurucz (1979) model
atmospheres, not corrected to empirical data, are 
in good agreement with
effective temperature derived from intensity-averaged $B-V$ colours using
the new Kurucz (1993) and the semi-empirical procedure defined above.
However, we caution that some of the original Baade-Wesselink analyses, from
which values in table 4 of CSJ were taken, adopted a semi-empirical calibration.

As a comparison, we also computed differences for the same set of stars with
the mean effective temperatures derived by McNamara (1997) using optical and
near infrared colours and the temperature calibrations given by the 
new Kurucz models. In Figure~\ref{fig6} it is easy to see that in this case,
although the average difference is not much larger than in the previous 
comparison 
($<$T$_{\rm eff, us} - T_{\rm eq, McNam}> = -101 \pm 16$ K (r.m.s. = 67~K, 
17 stars), a clear trend with temperature arises, likely due to the absence
of semi-empirical corrections to the models employed by McNamara (1997).

However, we can be confident that our temperature scale i) is fully homogeneous
for both field and cluster variables and ii) likely due to various, compensating
effects, the derived temperature are not much far away from the so-called 
equilibrium temperature of a pulsating star, as discussed in CSJ.

Note that in our analysis we avoid introducing any spurious results due to
the assumption that field and cluster variables have similar histories, as 
implicit in the CSJ and C98 analysis.

\begin{figure}
\vspace{8cm}
\includegraphics{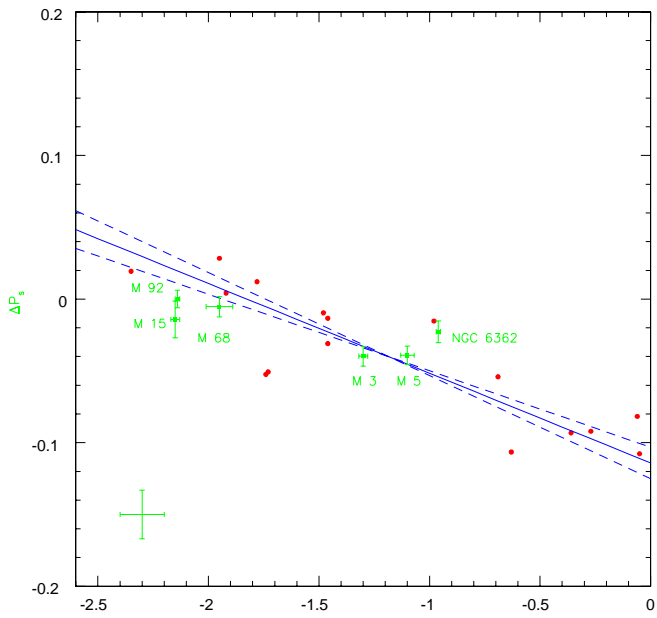}
\caption{ Run of the $\Delta$Ps $=$ dlogP$_{\rm oss - exp}$ as a function of
the metal abundance for stars in the Baade-Wesselink sample from CSJ (filled
circles). Mean $\Delta$Ps for cluster variables are displayed as filled squares,
with overimposed error bars corresponding to the internal errors. The linear best fit for
field stars is shown as a solid line which is the bisectors of direct and
inverse regressions (dashed lines). The errors bars plotted at the lower-left 
corner are those corresponding to systematic errors in the reddening and in the
metallicity scales, as discussed in the text. The metallicities are on
Carretta \& Gratton (1997) scale.}
\label{fig7} 
\end{figure}

As before we used M15 as a reference and derived a linear 
relationship 
as the best fit in the the logP--logT$_{eff}$ plane (a quadratic relation
is no longer justified). Entering in this relation with the 
derived effective
temperature for each stars, again we computed the difference
between the actually observed and the expected period of each star,
$\Delta$Ps $=$ dlogP$_{\rm oss - exp}$.
These differences are displayed in Figure~\ref{fig7} as a function of the
metal abundances, taking for each cluster the unweighted mean of all cluster
variables.
In order to evaluate the relevance of systematic errors in the reddening 
scale (which affects the derived temperatures $via$ $B-V$ colours) we 
repeated our analysis changing by $\pm 0.02$ the adopted reddening values for the cluster
stars. The corresponding 
uncertainty in the derived 
$\Delta$Ps $=$ dlogP$_{\rm oss - exp}$ would be about 0.02. 

As for systematic errors in the metallicity scale, the overall uncertainty
cannot be reduced below 0.1 dex, as discussed by Carretta et al. (2000). We
therefore adopted this figure as the possible systematic errors due to the
adoption of CG97 scale.

 A linear best fit
regression through the data of the 16 B-W field stars in Figure~\ref{fig7} 
gives :
\begin{equation}
\Delta Ps = -0.0625 {\rm [Fe/H]} -0.1141,
\end{equation}
which is the bisector of direct and inverse relations, 16 stars, 
correlation coefficient of $r=0.86$.
Nothing new in this relation, which slope is very similar to others 
obtained using field variables (e.g. Fernley 1993: -0.073).

On average, the cluster RR Lyraes
are not too much far away from the mean locus defined by field variables 
in Figure~\ref{fig7}. 
The impression by eye is of a hint for the RR Lyraes in clusters having 
slightly shorter periods than field
variables of similar metallicity.

We then derived the same fit also for the average
values defined by the cluster variables. Taking again the bisecant of relations
obtained exchanging dependent and independent variables, we derived from the
6 clusters considered here:
\begin{equation}
\Delta Ps = -0.0319 {\rm [Fe/H]} -0.0712,
\end{equation}
with a correlation coefficient $r=0.80$.

This slope is much less than the value derived
by the latest study of Sandage (1993). However, since we are mainly interested 
to obtain
a figure for the actual differences between the distribution of field and
cluster variables at fixed metallicity, 
we have computed the offset from each field RR Lyrae with respect to
the fit defined by the cluster variables.

The unweighted average derived from all the 16 field stars is now:
\begin{eqnarray}
<\delta(\Delta Ps)> & = & <\delta(dlogP_{star} - dlogP_{fit,GC})> \nonumber \\
                    & = & -0.0067\pm 0.0069.
\end{eqnarray}

On the other side, when considering only field stars with [Fe/H] values in the 
metallicity range spanned by globular clusters (i.e. excluding from the
average those more metal-rich than [Fe/H]$>-0.96$) the above figure become
$<\delta(\Delta Ps)> = 0.0066\pm 0.0075$, based on 10 stars. 

Both these results and the attached error bars simply tell us that any 
difference between field and cluster variables has to be considered at best not
very significant. In the following, we will use the average from Eq. (5), but
the discussion would not change even using the other value.

\section{Discussion}

If we now combine the result obtained in the previous section with the 
pulsational equation of (e.g.) Van Albada \& Baker (1971) we can simply write:
\begin{equation}
\Delta logP = (-0.0067\pm 0.0069) = 0.84 \Delta logL -0.68 \Delta logM,
\end{equation}
where all differential quantities (periods, luminosities and masses) correspond to 
mean
{\it field $-$ cluster} and are read at fixed temperature (and metallicity).
Since, according to the pulsation equation the period depends on both 
luminosity and mass 
to understand the physical meaning of our results we can make two
different assumptions concerning the mass of pulsating variables on the 
horizontal branch.

First we suppose that cluster and field variables were formed with 
identical masses and share similar histories and properties 
independently of the enviroment, and that the present mass is, for example, 
about 0.6 M$_\odot$,
as the classical value for globular cluster HB stars.
 
Then from equation (6) one derives that 
$\Delta logL_{fi-cl} = -0.008\pm 0.008$, and, neglecting at first order the 
bolometric corrections, $\Delta$ log$M_V$(fi-cl) $= 0.02\pm 0.02$. This 
means that field variables are approximately as luminous as cluster 
variables of same metallicity. This value is only a twenty percent in 
magnitude of the effect ($\sim 0.1$ mag)
required to explain the discrepancy between different distance indicators.

Given the above result, if we want that field variables are about 
0.1 mag $fainter$ than cluster RR Lyraes, we must release
the assumption of equal masses. 
If so, simple computations allow to obtain the values listed in Table 1.

\begin{table}
\caption{Masses of field RR Lyraes for various differences in L$_{HB}$ between
cluster and field variables}
\begin{tabular}{cccc}
\hline\hline
$\Delta logM_V$ & $\Delta logL$ & $\Delta$ logM & M$_{fi}$ \\
    (fi-cl)     &    (fi-cl)    &    (fi-cl)    & (if M$_{cl} = 0.6 M_\odot$) \\
\hline 
         +0.0    &     +0.00    &    $+0.010\pm 0.010$  & 0.614 M$_\odot$ \\   
         +0.1    &   $-$0.04    &    $-0.040\pm 0.010$  & 0.548 M$_\odot$ \\   
         +0.2    &   $-$0.08    &    $-0.089\pm 0.010$  & 0.489 M$_\odot$ \\   
\hline
\end{tabular}
\end{table}

It is
then clear that if the discrepancy in distances
calibrated using cluster and field variables is due to an intrinsic 
difference in
luminosity of $\sim 0.1$ mag, 
between field and cluster HB stars, we should also postulate that
HB stars in the more sparse field are about 0.05 M$_\odot$ less massive
than their cluster counterparts of similar temperature.

\subsection{Comparison with Zero Age Horizontal Branch models}

How this prediction derived from purely pulsational properties 
compares with evolutionary models of Zero Age Horizontal Branch
(ZAHB) stars ? Is there a basic parameter (e.g. core mass or
abundance of helium Y) which variation within plausible ranges could result 
into a difference of about 0.05 M$_\odot$ ? 

The theory of stellar evolution
has by long time secured the notion that the enhancement of Y in a ZAHB model
results in a star populating the HB at bluer colours (i.e. warmer temperatures)
and at brighter luminosities than a model with similar structural parameters
but lower Y abundance.

Actually, we are not interested in $how$ a HB
star has gained a larger Y abundance, but mainly in the consequences that
such enhancement could have.
However, stars obviously arrive on the HB following a well 
defined evolutionary path, and it is known that a good candidate to give
larger Y abundance is for instance the presence of deep extra-mixing whose
onset is likely related to some non-standard mechanism (see e.g. Cavallo et 
al. 1998 and quoted references).
In this scenario, both bluer colours and brighter luminosities are due to
the higher level attained by the stars to the red giant tip, with consequent
enhanced mass loss and helium core mass. The net result would thus be 
a star
which locates on the blue part of the horizontal branch.
In order to find such high-Y "cluster" star at the same colour
of a field RR Lyrae, i.e. inside the instability strip, 
the star should also have been born with a
higher mass, so to spend a part of his helium-core burning
as a pulsating variable.

To have a more quantitative insight into this 
scenario, we 
need a set of ZAHB models which explore 
systematically the variations of each structural parameter (mass, core mass, 
Y abundance) while keeping all the others fixed. 
Unfortunately, most of published 
studies assume a typical set of parameters and then follow the global 
$evolution$ of the stars, making it difficult, or almost impossible, to 
discriminate
among different involved parameters, and the new models computed by
Sweigart (1997) where enhancement of Y is explicitly treated are unfortunately
still unpublished. 
The most complete set of classical ZAHB models available so far are those 
Sweigart \& Gross (1976: SG76). 
In this respect, SG76 models are the
tool we need; all subsequent improvements in the input physics are
of minor relevance, in the present context, since we are mainly interested
to study differential effects.

Results of interpolations in the SG76 models are listed in Table 2, where we
report the variations obtained in the total ZAHB mass varying the 
core mass and the Y abundance, at fixed metallicity and temperature. 
Figures
are derived holding also a constant difference of $\Delta logL = -0.04$, 
(i.e. $\sim 0.1 mag)$.
 
\begin{table}
\caption{Mass differences from ZAHB models of Sweigart \& Gross (1976)
when varying core mass and helium abundance Y and assuming a constant difference
$\Delta logL = -0.04$.}
\begin{tabular}{ccc}
\hline\hline
$\Delta M_{core}$ & $\Delta$Y & $\Delta$M/M$_\odot$ \\
    (fi-cl)     &    (fi-cl)    &    (fi-cl)     \\
\hline 
         $-$0.000  &   $-$0.020   &    $-$0.004 \\
         $-$0.006  &   $-$0.012   &    $-$0.011 \\
         $-$0.014  &      0.000   &    $-$0.019 \\
\hline
\end{tabular}
\end{table}

Values in Table 2 read as the changes in M$_c$ and Y
of a theoretical ZAHB cluster star in order to be i) brighter 
(by $\sim 0.1$ mag) and, at same time, ii) more massive than a theoretical 
field star.

It is easy to see that
\begin{enumerate}
\item if Y is enhanced at fixed core mass, the resulting enhancement in 
mass is negligible, if compared to the required 0.06 M$_\odot$. 
Moreover, the solution would be a little unpalatable, since it is well known
that a higher Y implies higher luminosities of the star at the giant tip, but
also an increase of the luminosities of the RR Lyraes, and hence in their
periods (Sweigart 1997), at odd with present results. On the other side, a
higher Y, if primordial, would result in lower luminosities at the red giant
tip and in a smaller core mass (Sweigart, Greggio \& Renzini 1990);
\item if both Y and M$_c$ are increased, a larger increase in the mass appears
\item finally, it seems that better results are obtained simply with
an increased core mass. In fact, if we assume that cluster variables start 
their HB evolution with core masses 0.014 M$_\odot$ larger than field
stars (e.g. due to enhanced internal rotation, maybe driven by the denser
environment of their formation and evolution) then it is possible for such
stars to reach a difference of about 0.02 M$_\odot$. However, even in this 
case  the mass enhancement is a factor 3 smaller than derived from the pulsational 
analysis.
\end{enumerate}

\subsection{Summary and conclusions}

In this paper we explore the suggestion by Gratton (1998) that the existence
of an intrinsic difference of about 0.1 mag between the luminosity of 
field and cluster HB stars
could be responsible of the disagreement found between 
distance calibrations ultimately based on Hipparcos
parallaxes.

Following an approach very similar in principle to that presented 
by Catelan (1998), we used photometric data of RR Lyrae 
variables in the field and in globular clusters to study their pulsational
properties. Quantitative estimates of the amount of  possible differences
between the two kind of variables are given.

The differential comparison of the resultant period distributions leads to 
the following conclusions:
\begin{enumerate}
\item repeated trials using the original samples of variables adopted by
C98 but different metallicity scales, or sets of light curve parameters,
highlight that inhomogeneities in the data sets or intrinsic (even small)
internal errors could blur a luminosity
difference between field and cluster stars as derived from differences in 
their
period distributions at fixed metallicity;
\item when a homogeneous metallicity scale for both field and cluster 
RR Lyraes is used, and consistent temperatures from optical colours and the
new Kurucz's model atmospheres (tied to empirical calibrations) are derived, 
more 
quantitative and stringent estimates are possible;
\item we confirm on the whole C98 findings: at fixed temperature and 
metal abundance the run of periods with [Fe/H] is essentially the same for
both field and cluster RR Lyrae stars, even if there could be a small hint
for cluster RR Lyrae having slightly shorter periods than field variables of
the same metallicity;
\item our best estimate from 16 field RR Lyraes (with accurate parameters
from Baade-Wesselink analyses) and 114 cluster RR Lyraes with recent and 
accurate
photometry is that on average field stars have a difference of 
$-0.0067\pm 0.0069$ in $\Delta$ logP with respect to cluster variables. 
The statistical significance of this effect is then very scarce;
\item when combined with
the classical pulsational equation by (e.g.) van Albada \& Baker (1971), 
this difference formally implies that at fixed temperature and metallicity 
either the mass or
the luminosity (or both) of field and cluster variables must be slightly 
different, however assuming that field and cluster RR Lyraes were 
born with
same mass would result in a difference of only about two hundreth of a 
magnitude.
On the other hand, in order to achieve a 0.1 mag difference in HB 
luminosity as suggested by Gratton (1998), the field variables should be
about  $\sim 0.05 M_\odot$ less massive than their cluster
counterparts at same temperature and metal abundance.
\end{enumerate}

Unfortunately, the determination of the mass of a star is still one of the
most difficult problems in the observational astrophysics.
In the case of pop. II pulsating stars, one could exploit the pulsation
theory in order to derive mass estimates from 
periods; however, the exact value of the masses (and their run as a 
function
of [Fe/H]) is still an unsettled and controversial issue.

Results from a simple explorations of SG76 ZAHB grids, discussed in 
Section 4.1,
show that a larger core mass in the HB phase could give the larger masses
required to explain, at least in part, the discrepancy in HB luminosity 
tentatively suggested to exist between field and cluster stars. From a 
theoretical point of view, a good candidate to give larger
masses for stars born and evolved in the cluster dense environment 
is an enhanced Y abundance, possibly due to 
extra-mixing phenomena likely related to non-standard core rotation (see
Cavallo et al. 1998 and references therein).
Since evidences of deep mixing are only found in cluster giants and not
in field stars (Gratton et al. 2000), 
 the star birth-place and the density of 
the environment hosting its evolutionary life, in particular, could 
be responsible for the differences found between field and cluster objetcs.

However, according to Sweigart (1997), non-canonical helium-mixed models would
result into an increase of the luminosity of the RR Lyrae variables, and hence 
in
their pulsational periods.
This is clearly at odd with C98 and our results, which find  
no or very little differences between the period distributions of field and
cluster variables at fixed temperature.

\bigskip\bigskip\noindent
ACKNOWLEDGEMENTS

We warmly thank Marcio Catelan for kindly providing his original data
and a preprint of his paper in advance of publication. We also thank Alistair
Walker for sending us mean colour and parameters for variables in NGC 6362 in
machine ready form in advance of publication. It is a pleasure to thank 
Bernardo Salasnich and Leo Girardi for helpful discussions, as well the referee
for her/his useful comments.

\end{document}